# Premixed-Flame Shapes and Polynomials.

Bruno Denet[1*] and Guy Joulin[2#]

[1] Aix-Marseille Univ., IRPHE, UMR 7342 CNRS, Centrale Marseille,
Technopole de Château-Gombert, 49 rue Joliot-Curie, 13384 Marseille Cedex 13, France.

[2] Institut P-prime, UPR 3346 CNRS, ENSMA, Université de Poitiers,
1 rue Clément Ader, B.P. 40109, 86961 Futuroscope Cedex, Poitiers, France.

**Abstract**.

The nonlinear nonlocal Michelson-Sivashinsky equation for isolated crests of unstable flames is studied, using pole-decompositions as starting point. Polynomials encoding the numerically computed 2*N* flame-slope poles, and auxiliary ones, are found to closely follow a Meixner-Pollaczek recurrence; accurate steady crest shapes ensue for *N*≥3. Squeezed crests ruled by a discretized Burgers equation involve the same polynomials. Such explicit approximate shape still lack for finite-*N* pole-decomposed periodic flames, despite another empirical recurrence.

## 1. Introduction.

Sivashinsky [1] derived the first nonlinear evolution equation for the amplitude $\phi(t,x)$ of wrinkling of premixed gaseous flames subject to the nonlocal Darrieus [2]-Landau [3] (DL) hydrodynamic instability, when the Attwood number $\mathcal{A}=(E-1)/(E+1)$ based on the fresh-to-burnt density ratio $E>1$ is small. The evolution equation, first studied numerically in [4], is:

$$\phi_t + \tfrac{1}{2}\phi_x^2 - \nu\phi_{xx} + \mathcal{H}\{\phi_x\} = 0 \qquad (1)$$

once put in scaled form, and is often termed the Michelson-Sivashinsky (MS) equation. The subscripts in there denote partial derivatives in scaled time ($t$) and abscissa ($x$). The constant Markstein (dimensionless-) 'length' $\nu > 0$ controls the curvature-induced $\sim \nu\mathcal{A}^2\phi_{xx}$ changes in local speed (relative to fresh gas) of a flame front element [5]. The Hilbert transform is

$$\mathcal{H}\{\phi_x\}(x) = \frac{1}{\pi} \fint_{-\infty}^{+\infty} \frac{\phi_x(t,x')}{(x-x')}dx' \qquad (2)$$

if written in a form suitable for the isolated crests ($\phi_x(t,\pm\infty)=0$) to be first studied here; its form adapted to $2\pi$-periodic patterns, to which (1) also applies, will be recalled in **§5**. This nonlocal term $\mathcal{H}\{\phi_x\}$ encodes the DL hydrodynamic flame instability [1-3]: since $\mathcal{H}(e^{i\kappa x}) =$

---

[*] bruno.denet@irphe.univ-mrs.fr
[#] guy.joulin@lcd.ensma.fr:    Corresponding author, Tel(33)(0)549498186, Fax(33)(0)549498176

$i.\text{sgn}(\kappa)e^{i\kappa x}$ the growth/ decay rate of normal modes $\phi \sim e^{i\kappa x+\varpi t}$ of the linearized (1) indeed reads $\varpi = |\kappa| - \nu\kappa^2$. The nonlinear contribution $-\frac{1}{2}\phi_x^2$ to $\phi_t$ mainly [6] is of geometric origin [1]: the normal to the front locally makes a finite angle $\gamma \sim -\mathcal{A}\phi_x$ to the mean propagation direction (normal to $x$-axis), and $\cos(\gamma) - 1 \sim -\mathcal{A}^2\phi_x^2/2$.

Although its dimensioned version was originally derived as a leading order result for $\mathcal{A} \to 0^+$ [1], the MS equation (1) still rules flame dynamics if corrections (removed from (1) by re-scaling) due to two more orders are retained in the $\mathcal{A}$-expansion [6,7] to improve accuracy. Beside gaseous combustion, the MS equation governs other unstable fronts coupled to Laplacian fields: in doped semi-conductors [8] or in reactive infiltration [9].

Equation (1) exhibits a number of remarkable features, most notably the existence of pole-decompositions whereby the search for $\phi(t,x)$ is converted to a $2N$-body problem for the complex poles of the front slope [10,11]; see **§2**. Thanks to this property one can: (i) Explain the formation of front arches joined by sharper crests whose mergers ultimately produce the widest admissible steady cell; (ii) Access the latter's arc-length *vs.* wavelength curve [12], which yields the effective flame speed; (iii) Solve stability issues [13,14] without the effect of spurious noises hampering the non-self-adjoint linearized dynamics [15]; (iv) Compute pole density and front shapes for isolated crests, and then for periodic cells [11,16], if $N \gg 1$; (v) Study stretched crests [17]; (vi) Set up tools to study extensions of (1) that incorporate higher orders of the $\mathcal{A} \ll 1$ expansion, at least in the large-$N$ limit ([18,19] and Refs. therein).

It would be valuable to extend the latter works to *finite* pole numbers $2N$, as to encompass wrinkles of moderate wavelengths/amplitudes. New tools are needed and, just like in cases (iv)-(vi) above, studying solutions representing isolated front crests (localized bumps with $\phi_x(t,\pm\infty) = 0$, $\phi(t,-x) = \phi(t,x)$) might be a key step to take up first. The present numerical exploratory approach finds that crest-type solutions of (1) have intriguing – though as yet unexplained – relationships with known polynomials; this will ultimately provide one with approximate crest shapes in closed form that are accurate even for finite $N$ s.

## 2. Pole-decomposition for isolated crests

As shown in [10, 11] (1) admits solutions that have $\phi(t,x) = -2\nu\Sigma_{|k|=1,\ldots N} \ln(x - z_k(t))$ whence:

$$\phi_x(t,x) = \sum_{k=-N}^{N} \frac{-2\nu}{x - z_k(t)}, \qquad (3)$$

provided the complex-conjugate pairs of poles $z_k = \bar{z}_{-k}$, $|k| = 1,....N$, of the analytically-continued flame slope $\phi_x(t,z)$ in $z = x + iB$ plane satisfy :

$$\frac{dz_k}{dt} = \sum_{j=-N, j\neq k}^{+N} \frac{2\nu}{z_j - z_k} - i.\text{sgn}(\Im(z_k)) \ , \tag{4}$$

where $\Im(.)$ is an imaginary part. The basic identity $(x-z_k)^{-1}(x-z_j)^{-1} = (x-z_k)^{-1}(z_k - z_j)^{-1} + (x-z_j)^{-1}(z_j - z_k)^{-1}$ indeed allows one to transform the cross-terms generated when squaring (3), and a similar one followed by contour integrations in $z$-plane leads to $\mathcal{H}[1/(x-z_k)] = -i.\text{sgn}(\Im(z_k))/(x-z_k)$. Combined with $\phi_t$ this transform (1) in a sum of $C_k/(x-z_k)$ pieces, and (4) are the conditions $C_k = 0$ for (1) to be satisfied. The solutions (3) are localised, $\phi_x(t, |x|/\nu \to \infty) \approx -4N\nu/x$, and represent a collection of elementary isolated flame front crests, each associated with by a pair $z_k, z_{-k}$. As implied by (3) $e^{-\phi(t,z)/2\nu}$ is a polynomial with the $z_n(t)$ as simple zeros; its degree, the total number $2N$ of poles, is conserved yet arbitrary.

It is a known consequence [11] of the $(z_k - z_j)^{-1}$ interaction terms in (4) that nearby poles undergo mutual vertical repulsion and horizontal attraction. This mechanism ultimately produces a single steady arrangement of vertically aligned poles, say located at $z_k = iB_k = -iB_{-k}$, $1 \leq |k| \leq N$. Such time-independent crest shapes $\phi(x)$ obey steady versions of (1)-(4):

$$\tfrac{1}{2}\phi_x^2 + \mathcal{H}\{\phi_x\} - \nu\phi_{xx} = 0 \ , \tag{5}$$

$$\phi_x(x) = -\sum_{k=-N}^{N} \frac{2\nu}{x - iB_k} \ , \tag{6}$$

$$\sum_{j=-N, j\neq k}^{N} \frac{2\nu}{B_k - B_j} = \text{sgn}(B_k) \ , \tag{7}$$

$$\phi(x) = -2\nu \ln(i^{2N} P_{2N}(x/i)) + const. , \tag{8}$$

where the even monic polynomial $P_{2N}(B) = B^{2N} + ...$ has the real pole 'altitudes' $B_k$ as its zeros. The (dimensionless) Markstein [5] 'length' $\nu$ could obviously be scaled out from (7), but it is kept as is for future comparisons with other lengths beside the various $B_k$.

The $N = 1$ crest involves two poles $iB_{2,k}$, with $B_{2,\pm 1} = \pm \nu$ obtained from (7) as zeros of :

$$P_2(B) \equiv B^2 - \nu^2 \ . \tag{9}$$

For $N = 2$ the pole altitudes $B_{4,k} = \pm 3\nu(1 \pm 1/\sqrt{2})$ are the zeros of [11]:

$$P_4(B) \equiv B^4 - 27\nu^2 B^2 + \tfrac{81}{4}\nu^4 . \qquad (10)$$

When $N = 3$ a tedious algebra is already required to write the polynomial with irrational coefficients $P_6(B)$ [11], not to mention formulae for its zeros. And $P_{2N}(B)$ could not even be accessed analytically for $N > 3$; yet the zeros $B_{2N,k}$, $|k| = 1,...,N$, can be obtained numerically for any $N$, by Newton iterative resolution of (7) or as attractors of (4).

## 3. MS equation *vs*. MP polynomials

From the $B_{2N,k}, |k| = 1,...N$, computed on solving (7) for increasing $N$ s, monic polynomials $P_M(B) = B^M + ...$ are next defined by $P_0(B) = 1$, $P_1(B) = B$ and by (11) below for $M = 2, 3,...$:

$$P_{2N}(B) \equiv \prod_{k=1,..N} (B^2 - B_{2N,k}^2) ,$$

$$P_{2N+1}(B) \equiv B \prod_{k=1,..N} (B^2 - b_{2N,k}^2) . \qquad (11)$$

The auxiliary $b_k$-zeros in there obey equations similar to (7), yet with an extra $b_0 = 0$ whose charge is $-2\nu$, and for $|k| = 1,...,N$ are roots of conditions of electrostatic-like equilibriums:

$$\frac{2\nu}{b_k - 0} + \sum_{j=-N, j \neq k}^{N} \frac{2\nu}{b_k - b_j} = \text{sgn}(b_k); \qquad (12)$$

the equilibrium of $b_0 = 0$ is guaranteed by $b_{-k} = -b_k$. Such auxiliary $ib_{|k|\geq 1} \neq 0$ are poles of a smooth slope profile $F_x(x)$ governed by $-\tfrac{2\nu}{x} F_x + \tfrac{1}{2} F_x^2 + \mathcal{H}(F_x) - \nu F_{xx} = 0$ instead of (5); with $b_0 = 0$ included, the $ib_k$ are poles of a singular slope $f_x(x) := F_x(x) - \tfrac{2\nu}{x}$ obeying a locally-forced steady MS equation, *viz*. (5) with $2\pi\nu\delta(x)$ added to the right-hand side. The $b_{2N,k}$ were also computed numerically from (12), for increasing values of $N > 2$.

With the above convention on labels, two successive polynomials in the $\{P_M\}_{M=1,2...}$ sequence have opposite parities, $P_M(-B) = (-1)^M P_M(B)$. From the numerically built sequence one next performs the Euclidean division of each $P_M(B)$ by its antecedent $P_{M-1}(B)$, to produce:

$$P_M(B) = B P_{M-1}(B) - C(M) R_{M-2}(B) , \qquad (13)$$

where each polynomial remainder $R_{M-2}(B)$ has degree $M - 2$, parity $(-1)^{M-2}$, and may also be assumed monic on defining the coefficient $C(M)$ accordingly.

The numerically computed zeros $\beta_{M-2,k}$ of $R_{M-2}(B)$ happened to be real for all $M$, which was not a given. And, surprisingly enough, the roots $\beta_{2N-2,k}$ of $R_{2N-2}(B)=0$ were found to nearly coincide with the zeros, $B_{2N-2,k}$, of the 'Sivashinsky polynomials' $P_{2N-2}(B)$: the fractional differences are less than 0.2% for $N$ as low as 11, see Table I, and in all cases look smaller than typical $\mathcal{O}(1/N)$ quantities. Likewise, the auxiliary poles $b_{2N-1,k}$ were found to sit very close to the roots of $R_{2N-1}(B)=0$.

| $k =$ | 1 | 2 | 3 | 4 | 5 |
|---|---|---|---|---|---|
| $\beta_{20,k}/\pi v$ | .2164 | .9396 | 1.859 | 2.960 | 4.250 |
| $B_{20,k}/\pi v$ | .2156 | .9383 | 1.857 | 2.958 | 4.247 |
| $k =$ | 6 | 7 | 8 | 9 | 10 |
| $\beta_{20,k}/\pi v$ | 5.749 | 7.498 | 9.562 | 12.08 | 15.40 |
| $B_{20,k}/\pi v$ | 5.746 | 7.493 | 9.552 | 12.07 | 15.39 |

Table I : Comparison of the positive roots $\beta_{20,k}$ of the monic remainder $R_{20}(B)$ introduced in eq.(13) with those, $B_{20,k}$, of the Sivashinsky polynomial $P_{20}(B)$ defined by eqs.(7)(11).

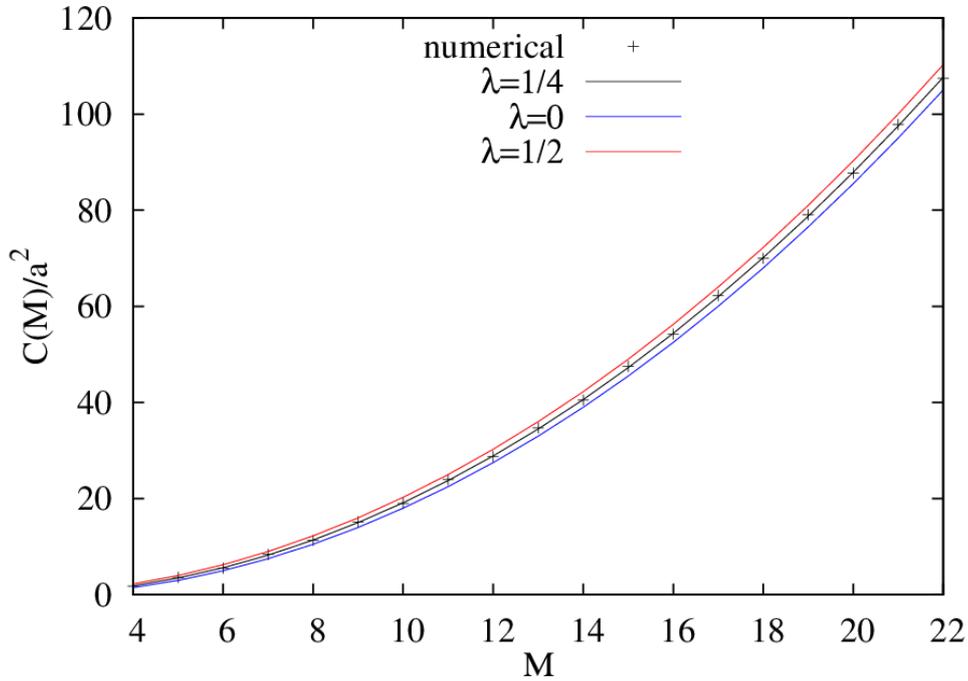

FIG.1 Scaled coefficients $C(M)/a^2$ vs. polynomial index $M$ with $a = \pi v$ from (19). Solid lines: $(M-1)(M+2\lambda-2)/4$ for $\lambda = 1/4$ (middle, black), $\lambda = 0$ (lowest, blue) or $\lambda = 1/2$ (upper, red); Symbols (+) : $C(M)/a^2$ as numerically determined from the polynomials $P_M(B)$ and equation (13).

The coefficient $C(M)$ in (13) is necessarily proportional to $\nu^2$ on 'dimensional' grounds. Furthermore $C(M)$ is numerically found to accurately conform to $a^2 \frac{1}{4}(M-1)(M-\frac{3}{2})$, with $a = \pi\nu$ as suggested in Eq.(19) below: see Fig.1. In other words, the flame-crest related $\{P_M(B)\}_{M=2,3...}$ sequence nearly follows the same recurrence as the monic symmetric Meixner-Pollaczek (MP) polynomials $P_M^{(1/4)}(\xi)$ [20, 21], up to the uniform re-scaling $B \leftrightarrow \xi = B/a$: the monic symmetric MP polynomials $P_M^{(\lambda)}(\xi)$ are indeed defined[1] for $M = 1, 2, ...$ and any real $\lambda$ by the 3-term recurrence

$$P_M^{(\lambda)}(\xi) = \xi P_{M-1}^{(\lambda)}(\xi) - \tfrac{1}{4}(M-1)(M+2\lambda-2) P_{M-2}^{(\lambda)}(\xi) , \qquad (14)$$

with $P_{-1}^{(\lambda)}(\xi) \equiv 0$, $P_0^{(\lambda)}(\xi) \equiv 1$ as initial conditions, whence $P_1^{(\lambda)}(\xi) = \xi$. For $\lambda > 0$ these are orthogonal over $-\infty < \xi < \infty$ with respect to the weight function $W^{(\lambda)}(\xi) = |\Gamma(\lambda + i\xi)|^2$, where $\Gamma(.)$ is the gamma-function, are of degree $M$ and have the definite parity $(-1)^M$. Their $M$ zeros $\xi_{M,k} = -\xi_{M,-k}$ are all real, $\xi_{M,0} = 0$ being admissible only if $M$ is odd. And, contrary to the Sivashinsky polynomials $P_{2N}(B)$, the $P_M^{(\lambda)}(\xi)$ are all known in *explicit* form as $P_M^{(\lambda)}(\xi) \sim i^M {}_2F_1(-M, \lambda + i\xi, 2\lambda \,|\, 2)$ [21], where ${}_2F_1(a,b,c\,|\,z) = \Sigma_{n\geq 0} (h_n z^n)$ with $h_0 = 1$ and $h_{n+1}/h_n = (a+n)(b+n)/(c+n)(n+1)$ is the usual (yet here terminating) Hypergeometric series [21].

Moreover, each $P_M^{(\lambda)}(\xi)$ satisfies the linear *difference* equation [20, 21]:

$$(\lambda - i\xi) P_M(\xi+i) - 2(M+\lambda) P_M(\xi) + (\lambda + i\xi) P_M(\xi - i) = 0 . \qquad (15)$$

The $\xi_{2N,k}$ thus obey $(\lambda - i\xi_{2N,k}) P_{2N}^{(\lambda)}(\xi_{2N,k} + i) + (\lambda + i\xi_{2N,k}) P_{2N}^{(\lambda)}(\xi_{2N,k} - i) = 0$ and, since $P_{2N}^{(\lambda)}(\xi) \equiv (\xi - \xi_{2N,-N})...(\xi - \xi_{2N,N})$, they are roots of Bethe-Ansatz (BA) type [22] equations:

$$\prod_{j=-N, j\neq k}^{+N} \frac{\xi_k - \xi_j + i}{\xi_k - \xi_j - i} = \frac{1 + i\xi_k/\lambda}{1 - i\xi_k/\lambda}, \quad |k| = 1, ..., N . \qquad (16)$$

Taking logarithms and defining $A(s) := \tan^{-1}(s)$, the crest-slope poles should then have:

$$\sum_{j=-N, j\neq k}^{+N} A(a/(B_k - B_j)) = A(B_k/a\lambda) , \qquad (17)$$

---

[1] The Secondary polynomials $P_{*M}^{(\lambda)}(\xi) \propto \int_{-\infty}^{+\infty} (P_M^{(\lambda)}(\xi) - P_M^{(\lambda)}(\zeta)) W^{(\lambda)}(\zeta) d\zeta/(\xi - \zeta)$, $P_{*0}^{(\lambda)}(\xi) = 0$, $P_{*1}^{(\lambda)}(\xi) = 1$, $P_{*2}^{(\lambda)}(\xi) = \xi$ ... etc. also are [20], but do not provide separate good fits when $\lambda = \tfrac{1}{4}$; *e.g.*, $P_{*3}^{(1/4)}(\xi) = \xi^2 - \tfrac{3}{4}$.

$|k|=1,...,N$, if the correspondence $B_{2N,k} = a\xi_{2N,k}$ was indeed exact. At large enough $N$ the $B_k$s scale as $N\nu >> a \sim \nu$ [16]; in this scaling limit (17) would simplify to:

$$\sum_{j=-N, j\neq k}^{+N} \frac{a}{B_k - B_j} \underset{B_k>>a}{\approx} \frac{\pi}{2}\text{sgn}(B_k), \qquad (18)$$

which can be made coincide with (7), whatever $\lambda$ and the (large enough) $N$ are, on selecting:

$$a = \pi\nu. \qquad (19)$$

This choice was already made when drawing Fig.1. It ensures that the $B_{2N,k}$s and the scaled zeros $a\xi_{2N,k}$ are distributed for $N \to \infty$ according to the same Nevai-Ullman type density $\rho(B) = \frac{1}{\pi^2 \nu}\cosh^{-1}(2N\pi\nu/|B|)$ [16, 23] whatever the value of $\lambda$ is: it is indeed known [23] that $\lambda$ mainly encodes information about the lowest zeros, whence it disappeared from (18).

The correspondence between zeros of MP polynomials and poles of MS flame slope is not exact though. For example (17)(19) with $\lambda = \frac{1}{4}$ predict $\pi\xi_{2,1} = \pi(\frac{1}{2}\lambda)^{1/2} \approx 1.11$ when $N=1$, whereas (9) yields $B_{2,1}/\nu = 1$; yet the difference already is only 11% (it would vanish if $\lambda$ was $2/\pi^2 \approx 0.203$). For $N=2$, the solutions of (17)(19) determined with $\lambda = \frac{1}{4}$ differ from those of (10) by less than 7% on $B_{4,1}$ and 0.1% on $B_{4,2}$.

| $k =$ | 1 | 2 | 3 | 4 | 5 |
|---|---|---|---|---|---|
| $B_{20,k}/\pi\nu$ | .2156 | .9383 | 1.857 | 2.958 | 4.247 |
| $P_{20}^{(1/4)}$ roots | .2178 | .9018 | 1.820 | 2.924 | 4.219 |
| $P_{20}^{(2/\pi^2)}$ roots | .1995 | .8787 | 1.795 | 2.897 | 4.189 |
| $k =$ | 6 | 7 | 8 | 9 | 10 |
| $B_{20,k}/\pi\nu$ | 5.746 | 7.493 | 9.558 | 12.07 | 15.39 |
| $P_{20}^{(1/4)}$ roots | 5.724 | 7.479 | 9.552 | 12.08 | 15.40 |
| $P_{20}^{(2/\pi^2)}$ roots | 5.691 | 7.444 | 9.514 | 12.04 | 15.36 |

Table II: The MS zeros of $P_{20}(B)$, Eqs.(7)(11), compared with those of $P_{20}^{(\lambda)}(B/\pi\nu)$ for $\lambda = \frac{1}{4}$ or $\lambda = 2/\pi^2$.

Table II next compares the scaled poles locations $B_{2N,k}/\pi\nu$, numerically obtained from (7) for $N=10$, with the roots $\xi_{20,k}$ of $P_{20}^{(\lambda=1/4)}(\xi) = 0$ computed from the BA-like equations (17) and (19): even with $N$ this low the fractional difference at most is $\approx 3\%$ at $k = \pm 2$, and it decays very rapidly as $|k|$ grows. That the Sivashinsky and the MP polynomials share the same density of zeros at $|B| \geq \mathcal{O}(\nu)$ [11, 23] when $N \to \infty$ (hence yield the same crest shapes at $|x| \geq \mathcal{O}(\nu)$) likely favoured a good fit, but a uniformly good accuracy could not be reached without selecting $\lambda \approx \frac{1}{4}$: see Table II for $\lambda = 2/\pi^2$ or Fig.1 with $\lambda = 0$ or $\frac{1}{2}$. As expected, changing $\lambda$ mainly affects the lowest zeros; yet $P_{20}^{(\lambda=1/2)}(\xi)$, a function with special properties as regards $\text{sgn}(\xi)$ [23] and Hilbert transform [18, 23], already has all its roots significantly off the $B_{20,k}/\pi\nu$ (e.g., $\xi_{20,1} \approx 0.29$, $\xi_{20,5} \approx 4.37$, $\xi_{20,10} \approx 15.64$).

The only known [11] exact consequence of (7), $\Sigma_{k=1}^{k=N} B_{2N,k}/\nu = N(2N-1)$, is dominated by the more distant poles, whence $\pi \Sigma_{k=1}^{k=N} \zeta_{2N,k}$ retrieves it accurately, e.g., 14.994... for $N = 3$.

To complete this section one may recall that locating the poles $iB_k$ was just one means to access the flame-crest slope $\phi_x(x)$; but once the MP fit with $\lambda = \frac{1}{4}$ is adopted, the explicitly known $P_{2N}^{(1/4)}(x/i\pi\nu)$ gives the approximate crest shape itself in closed form:

$$\phi(x) \approx -2\nu \ln[_2F_1(-2N,\ \tfrac{1}{4} + x/\pi\nu,\ \tfrac{1}{2}|\ 2)] + const. \qquad (20)$$

As is illustrated in Fig.2, (20) already is pretty close to the exact MS crest shape for $N = 3$; all $N \geq 4$ would give coincident approximate and exact curves at this scale ($|x| \geq \mathcal{O}(\nu)$).

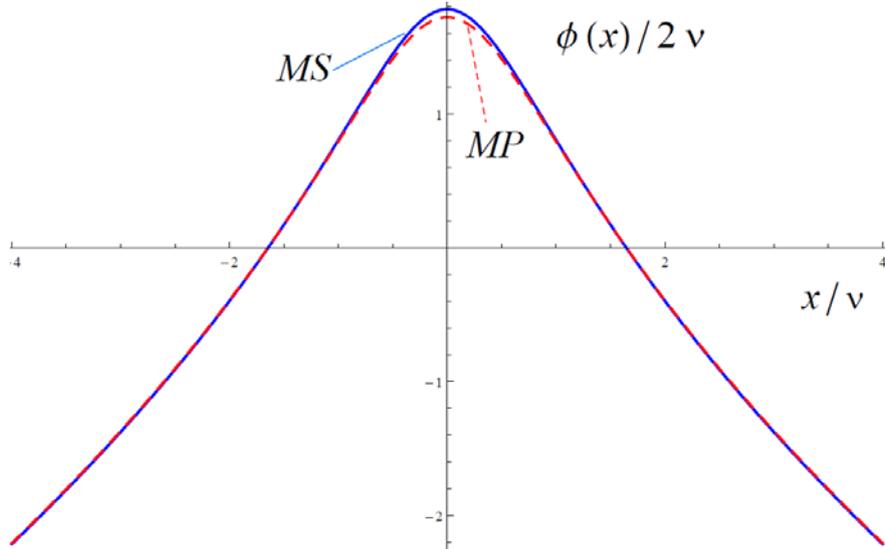

FIG.2 Crest shapes for $N = 3$. Solid blue line: MS shape from (7)(8); dashed red line: approximation (20).

One can try to improve (20) by invoking more general polynomials, e.g., the Symmetric Continuous Hahn polynomials $p_M^{(\lambda_1,\lambda_2)}(\zeta)$ that lie one 'shelf' above the $P_M^{(\lambda)}(\xi)$ in the Askey

Scheme [20, 21]. Those of interest here have[2] $\lambda_j > 0$, are orthogonal over $-\infty < \zeta < \infty$ with respect to $|\Gamma(\tfrac{1}{2}\lambda_1 + i\zeta)\Gamma(\tfrac{1}{2}\lambda_2 + i\zeta)|^2$, have parities $(-1)^M$ and also satisfy a 3-point difference equation [20]. The BA equations for the roots $\zeta_{2N,k}$ of $p_{2N}^{(\lambda_1,\lambda_2)}(\zeta)$ resemble (17) except for $A(2\zeta_{2N,k}/\lambda_1) + A(2\zeta_{2N,k}/\lambda_2)$ in the right-hand sides [20, 21]: because the latter $\tan^{-1}(.)$ functions add up to $\pi\,\mathrm{sgn}(\zeta_{2N,k})$ at large scale, the correspondence between $B$ and $\zeta = B/c$ has $c = 2\pi\nu$ instead of (19), i.e. $\xi_{2N,k} \leftrightarrow B_{2N,k}/\pi\nu \leftrightarrow 2\zeta_{2N,k}$, and the $c\zeta_{2N,k} \sim N\nu$ acquire the previous density $\rho(B)$ for $N \to \infty$. The Hahn family $\{p_M^{(\lambda_1,\lambda_2)}(\tfrac{1}{2}\zeta)\}_{M=1,2...}$ includes the MP sequence, as the Legendre identity $\Gamma(u)\Gamma(u + \tfrac{1}{2}) = \sqrt{\pi}\,2^{1-2u}\Gamma(2u)$ implies $p_M^{(\lambda,\lambda+1)}(\zeta) \propto P_M^{(\lambda)}(2\zeta)$ [25]. Optimizing $(\lambda_1,\lambda_2)$ around $(\lambda,\lambda+1)$ could thus make the $2\pi\nu\zeta_{2N,k}$ fit the lowest poles $B_{2N,k}$ even better than the roots $\pi\nu\xi_{2N,k}$ did; unexpectedly, the enhanced flexibility did not permit any significant (and uniform) improvement over (20).

### 4. Discretizing a local equation.

MP polynomials provide one with accurate steady localised solutions of the MS equation as soon as $N$ is moderately large $(>3)$; it is then natural to inquire whether close relatives of (5) itself could have solutions that are even better described by the MP sequence.

In this context consider the new, 'Toda-style' nonlocal equation:

$$\tfrac{\nu}{a^2}[e^{(\phi(x)-\phi(x+a))/2\nu} - 2 + e^{(\phi(x)-\phi(x-a))/2\nu}]$$
$$= -\tfrac{\sigma}{2a}x.\{e^{(\phi(x)-\phi(x+a))/2\nu} - e^{(\phi(x)-\phi(x-a))/2\nu}\} + \tfrac{U}{2\nu} \quad . \tag{21}$$

If $\phi(x)$ were to vary only slowly over the $a$ length-scale, Taylor-expanding $\phi(x) - \phi(x \pm a)$ as $\mp a\phi_x(x) - \tfrac{1}{2}a^2\phi_{xx}(x) + ...$ would make (21) resume a local, Burgers type equation:

$$-U + \tfrac{1}{2}\phi_x^2 - \sigma x\phi_x - \nu\phi_{xx} = 0 \quad , \tag{22}$$

where a positive reciprocal 'length' $\sigma$ tends to squeeze the crest [17], or else stretches it; $U \neq 0$ in (22) would be the speed of a propagating crest $\phi(t,x) = -Ut + \phi(x)$. Equation (22) admits a pole-decomposition of same form as (6), as is best shown on invoking the Hopf-Cole trick $\phi(x) = -2\nu\ln Q(x)$ to transform (22) into a linear problem where $U$ is an eigenvalue:

---

[2] $\lambda_j = \lambda' \pm i\lambda''$, $\lambda' > 0$, could have been envisaged as well, but this does not generalise the MP case.

$$\nu \frac{Q_{xx}}{Q} = -\sigma x . \frac{Q_x}{Q} + \frac{U}{2\nu} \ . \tag{23}$$

The BA equations for the zeros of $Q(x)$, obtained on equating the residues of either side of (23), look like (7) yet have $\sigma B_k$ instead of $\text{sgn}(B_k)$: as is well known [21] such (new-) pole altitudes are zeros of a Hermite polynomial $H_{2N}(B\sqrt{\sigma/2\nu})$ provided $\sigma > 0$, $U/2\nu$ is $2\sigma N$ and the crest shape implied by (22) reads $-2\nu \ln[i^{2N} H_{2N}(ix\sqrt{\sigma/2\nu})] + \text{const.}$ instead of (20). In analogy with this, setting $\phi(x) = -2\nu \ln Q(x)$ in (21) produces:

$$\nu \frac{Q(x+a) - 2Q(x) + Q(x-a)}{a^2 Q(x)} = -\sigma x . \{\frac{Q(x+a) - Q(x-a)}{2a Q(x)}\} + \frac{U}{2\nu} \ , \tag{24}$$

which is nothing but a discretized form of the 'squeezed' Burgers equation (22) once it is Hopf-Cole transformed as in (23). The point is that (24) constitutes a mere rearrangement of the Meixner-Pollaczek difference equation (15), provided $Q(x) \propto P_M^{(\lambda)}(x/ia)$ and:

$$\lambda = 2\nu/(\sigma a^2) \ , \quad U/2\nu = \sigma M \ . \tag{25}$$

For $M = 2N$, *exact* solutions to (21) thus are $\phi(x) = -2\nu \ln[i^{2N} P_{2N}^{(2\nu/\sigma a^2)}(x/ia)] + \text{const.}$. Since the first bracket of (21) mirrors $\frac{1}{2}\phi_x^2 - \nu \phi_{xx}$ in (5), the terms on the right of (24) should somehow be analogous to $\mathcal{H}\{-\phi_x\}$ despite their explicit dependence on $x$. A qualitative analogy does exist because $-\mathcal{H}\{1/(x-iB_k)\} = i\,\text{sgn}(B_k)/(x-iB_k)$ and $\sigma.(x/(x-iB_k)-1) = i\sigma B_k/(x-iB_k)$ mainly differ for $\sigma > 0$ by the already evoked replacement of $\text{sgn}(B_k)$ by $\sigma B_k$, and the fraction on the right of (24) comprises $2N$ additive $1/(x-iB_k)$ pieces.

This does not explain the quantitative agreements though; nor the fact that (25) links $\lambda$ to $\sigma$ without singling out a particular $\lambda$ whereas $\lambda = \frac{1}{4}$ is needed in order for (20) to correctly approximate MS isolated crest shapes.

## 5. Discussion, open problems.

Since MP polynomials give access to exact solutions to (21) despite a right-hand side quantitatively different from $\mathcal{H}\{-\phi_x\}$, the approximate recurrence among the $P_M(B)$s, (20) and the results summarized in Fig.1 or Table II cannot be blithely 'explained' by equation (5) being close to exactly solvable. Why then $P_{M+1}(B) - BP_M(B)$ in (13) nearly has a single component if projected onto a $P_M^{(\lambda)}(B/\pi\nu)$ basis, why precisely this basis, and why $\lambda = \frac{1}{4}$?

Admittedly the $P_M(B)$s and *all* the MP polynomials share the same density of zeros if $N \gg 1$, $\tan^{-1}(Z) - Z \approx -Z^3/3$ quickly gets negligible in (17) as $Z = a/(B_k - B_j) \sim 1/N$ decays, and a small $\lambda$ makes the right-hand side of (17) resume $\frac{\pi}{2}\text{sgn}(B_k)$ as soon as $B_k/\nu \gg \lambda$. As such approximations fail if $B_k/a\lambda$ (*e.g.*, $B_1/a\lambda$) or $Z$ (*e.g.*, $a/2B_1$) get $\mathcal{O}(1)$, adjusting $\lambda$ could conceivably better predict the smallest poles. But why the same $\lambda = \frac{1}{4}$ makes it at once, Fig.1, and why a 3-term recurrence involving the auxiliary $b_k$ and (12) ensues, remains unexplained. Since $\Sigma_{j \neq k} f(B_k - B_j) = g(B_k)$ pole- or BA equations can all be re-written as $\partial \mathcal{E} / \partial B_k = 0$ in terms of suitable interaction energies $\mathcal{E}[\{B_k\}_{k=-N}^{k=+N}]$ [11], approximating pole altitudes by zeros of polynomial solutions of differential or/and difference equations could possibly be formulated as an optimization (*e.g.*, of $\mathcal{E}_{MS}$) with a constraint (*e.g.*, $\mathcal{E}_{MP} = const.$).

At any rate, it would be important for wrinkled-flame theory to understand and systematize the 'coincidences' leading to (20). A related problem of much interest indeed concerns the possibility of generalizing (20) as to approximate $2\pi$-periodic patterns. $\nu < 1$ then represents the neutral-to-actual wavelength ratio and can no longer be scaled out, while the Hilbert transform is conveniently redefined as a convolution (over $|x'| \leq \pi$) with $\frac{1}{2\pi}\cot(\frac{1}{2}(x-x'))$. The 'steady' patterns from the MS equation (1) now drift at constant speed $U$ and read $\phi(t, x) = -Ut + \phi(x)$. Up to an arbitrary additive constant the wavy front shape then is $\phi(x) = -2\nu \ln[\Sigma_{|k|=1,...N} \sin(\frac{1}{2}(x - iB_k))]$ whence (6) becomes $\phi_x = -\nu \Sigma_{|k|=1,...N} \cot(\frac{1}{2}(x - iB_k))$. The new flame-slope poles $iB_k$ are to be found for $|k| = 1,...,N$ on solving [11]:

$$\sum_{j=-N, j \neq k}^{+N} \nu \coth(\tfrac{1}{2}(B_k - B_j)) = \text{sgn}(B_k) \qquad (26)$$

instead of (7). Thanks to $\sin(a - ib)\sin(a + ib) \equiv \sin^2(a) + \sinh^2(b)$, $e^{-\phi/2\nu}$ becomes an even polynomial of $\sin(\frac{1}{2}x)$, with roots $\pm i \sinh(\frac{1}{2}B_k) := \pm iS_k$ numerically accessible from (26); and the same for auxiliary $b_k$'s from equations resembling (26) with $\nu \coth(\frac{1}{2}(b_k - 0))$ added to the left-hand side (see (12)), which yields the $s_k = \sinh(\frac{1}{2}b_k)$. Numerical polynomials $P_M(S)$ of $S := \sinh(\frac{1}{2}B)$ (whose roots are the $S_{2N,k}$, or $0$ and the $s_{2N,k}$) can be built just like in (11) and provide monic remainders $R_{M-2}(S)$ and coefficients $C(M,\nu)$ as in (13). At fixed large enough $1/\nu > 1$, the $R_{M-2}(S)$ happen to again accurately coincide with the corresponding $P_{M-2}(S)$, at least for not-too-large degrees $M$. The sequence of $P_M(S)$s approximately related by a recurrence indeed has a *finite* $\nu$-dependent length, as $\coth(\frac{1}{2}(B_{2N,N} - B_{2N,j})) \geq 1$ in (26) implies $(2N-1)\nu \leq 1$ then $N \leq N_{opt}(\nu) := \lfloor \frac{1}{2}(1+\frac{1}{\nu}) \rfloor$; and the same for the auxiliary altitudes $b_k$, since $b_{2N,N} \leq \infty$ necessitates $(2N+1-1)\nu \leq 1$. If $1/\nu$ approaches the even integer $2m$ from above, the *monic* polynomial $P_{2m+1}(S)$ has a growing $s_{2m,m}$ root while those of $P_{2m}(S)$, $P_{2m-1}(S)$ stay finite, whence the coefficient $C(2m+1,\nu)$ must diverge; likewise if $1/\nu \to (2n+1)^+$, $S_{2n,n}$ and $C(2n,\nu)$ diverge. Although the existence of a finite-length recurrence and the possibility of small denominators in $C(M,\nu)$ are reminiscent of Racah, Hahn or $q$-Hahn polynomials [20, 21], no adequate analog of (15) was found.

Yet a $2\pi$-periodic analogue of (22) does exist, *viz.*:

$$-U + \tfrac{1}{2}\phi_x^2 - 2\sigma \tan(\tfrac{1}{2}x)\phi_x = \nu \phi_{xx}, \qquad (27)$$

and yields pole equations resembling (26), now with $2\sigma \tanh(\frac{1}{2}B_k)$ as right-hand side. This new local equation (27) is solvable in terms of even Gegenbauer polynomials of $\sin(\frac{1}{2}x)$, just like (22) was in terms of Hermite's, and exhibits similarities with (1) specialized to $2\pi$-periodic patterns: if $2\sigma = 1$, (27) has the same $N_{opt}(\nu)$ and $U = 2\nu N(1-\nu N)$ [16] as from the MS equation (1); also, because $\text{sgn}(B)$ and $\tanh(\frac{1}{2}B)$ look alike at large scale, the slope poles then acquire the same density as the MS pattern for $N \sim 1/\nu \gg 1$ [16]. This again results from a qualitative analogy: here $\mathcal{H}\{\phi_x\} = 2N\nu + i\nu \Sigma_k \text{sgn}(B_k) \cot(\frac{1}{2}(x-iB_k))$ whereas the local equation (27) has $-\tan(\frac{1}{2}x)\phi_x = 2N\nu + i\nu \Sigma_k \tanh(\frac{1}{2}B_k) \cot(\frac{1}{2}(x-iB_k))$. However, although the analogies begin as in **§4** no discretization-based generalization of (20) could be identified as to approximate $2\pi$-periodic MS flame fronts when $N = \mathcal{O}(1)$; and the reason why $R_{2N-2}(S) \approx P_{2N-2}(S)$ for large enough $N \leq N_{opt}(\nu)$ also remains unexplained.

Such open problems will hopefully prompt more mathematical investigations. It is indeed important to decide whether the present findings constitute curiosities/coincidences – yet ones compatible with 3-term recurrences and known pole/zero densities – or point to yet unraveled properties of the MS equation that could help one construct other approximate flame shapes.

## 6. References.